\documentstyle[preprint,pra,floats,aps,psfig]{revtex}

\begin{document}

\title{Nuclear spin conversion in formaldehyde}
\author{P.L.~Chapovsky\thanks{E-mail: chapovsky@iae.nsk.su}}

\address{Institute of Automation and Electrometry,\\
          Russian Academy of Sciences, 630090 Novosibirsk, Russia}
         
\date{\today}

\draft
\maketitle

\begin{abstract}
Theoretical model of the nuclear spin conversion in formaldehyde
(H$_2$CO) has been developed. The conversion is governed by the
intramolecular spin-rotation mixing of molecular ortho and para states.
The rate of conversion has been found equal 
$\gamma/P = 1.4\cdot10^{-4}$~s$^{-1}$/Torr.
Temperature dependence of the spin conversion has been predicted 
to be weak in the wide temperature range $T=200-900$~K.
\end{abstract}

\vspace{2cm}

\section{Introduction}

It is well known that many symmetrical molecules exist in nature only in 
the form of nuclear spin isomers \cite{Landau81}. The spin isomers 
demonstrate anomalous stability. For example, the ortho and para 
isomers of H$_2$ survive almost 1 year at 1~atm and room 
temperature \cite{Farkas35}. Existence of spin isomers is well-understood 
in the framework of the spin-statistics relation in quantum mechanics. 
On the other hand, the dynamical part of the problem, viz., the isomer
stability, their conversion rates and responsible conversion mechanisms
are less clear. This is because experimental data on spin isomers are 
rare due to substantial difficulties in preparation of enriched spin isomer
samples. Thus each new experimental result in this field deserves close 
attention.

In this paper we perform theoretical analysis of the 
nuclear spin conversion in formaldehyde (H$_2$CO). This process was
considered theoretically previously \cite{Curl67JCP}. But it is
worthwhile to perform new analysis in order to account new 
important information. First, the formaldehyde spin conversion  
was recently investigated experimentally and the gas phase 
conversion rate was measured for the first time \cite{Peters99CPL}.  
Second, formaldehyde molecular structure and molecular 
spectroscopic parameters have been determined very accurately
\cite{Bocquet96JMS,Carter96JMS,Muller00JMS}. Third, simple theoretical 
model of nuclear spin conversion in asymmetric tops was 
developed and tested by the spin conversion in ethylene \cite{Chap00PRA}. 
All these circumstances allow to advance 
significantly theory of the spin conversion in formaldehyde.

\section{Direct and indirect conversion mechanisms}

There are two main mechanisms known for the gas phase spin 
conversion in molecules. The first one, {\it direct}, consists 
in the following. In the course of collision inhomogeneous magnetic 
field produced by collision partner induces direct transitions in 
the test molecule between spin states. This mechanism 
is responsible for the hydrogen isomer conversion by paramagnetic 
O$_{2}$ \cite{Wigner33ZfPC}. It is important for us  
that the hydrogen conversion  induced by O$_{2}$ (the rate is equal to
$8\cdot10^{-6}$~s$^{-1}$/Torr \cite{Farkas35}) appears to be 
much slower than the spin conversion in formaldehyde \cite{Peters99CPL}, 
\begin{equation}
     \gamma_{exp}/P = (1.1\pm0.3)\cdot10^{-3}~{\text{s}}^{-1}/{\text{Torr}}.
\label{peters}
\end{equation}
Note that we refer here to the formaldehyde ortho-para equilibration rate.
In notations of \cite{Peters99CPL} this rate is equal to $k_1+k_2$, where
$k_1$ and $k_2$ are the ortho-to-para and para-to-ortho conversion
rates, respectively.

Hydrogen is an exceptional molecule due to its anomalously large
ortho-para level spacing and high symmetry.
It is more appropriate for the present discussion to consider the 
direct mechanism in polyatomic molecules.
The isomer conversion in CH$_3$F induced by 
O$_2$ was investigated theoretically and experimentally
in \cite{Chap96CPL,Nagels98CPL}. 
It was found that this mechanism provides the conversion rate  
on the order of $10^{-4}-10^{-5}$~s$^{-1}$/Torr. These rates
refer to the collisions with paramagnetic O$_2$ which have magnetic
moment close to Bohr magneton, $\mu_B$. An experimental arrangement
in \cite{Peters99CPL} corresponds to ``nonmagnetic'' 
collision partner having a magnetic moment on the order of nuclear 
magneton, $\mu_n$, thus by $10^3$ times smaller.
The conversion rate by direct mechanism depends on magnetic moment as $\mu^2$
\cite{Chap96CPL}. It implies that the conversion by direct 
process is too slow to be taken into account for the
formaldehyde conversion. In fact, the authors 
\cite{Peters99CPL} had arrived at the same conclusion.

Second mechanism is applicable to molecules in which 
different spin states can be quantum mechanically mixed by
weak {\it intramolecular} perturbation. Mixing and interruption
of this mixing by collisions with the surrounding particles
result in the isomer conversion. We will refer to this 
mechanism as {\it quantum relaxation}. Quantum relaxation can 
provide significantly faster conversion than direct process does 
for nonmagnetic particles. It can be illustrated by a few examples in
which quantum relaxation was established to be a leading process.
The most studied case is the spin conversion in $^{13}$CH$_3$F
(for the review see \cite{Chap99ARPC}). The conversion rate in
this molecule was determined as $(12.2\pm0.6)\cdot10^{-3}$~s$^{-1}$/Torr.
The conversion rate in ethylene ($^{13}$CCH$_{4}$) is equal to 
$(5.2\pm0.8)\cdot10^{-4}$~s$^{-1}$/Torr \cite{Chap00CPL}. 
In this molecule
too the quantum relaxation is the leading process \cite{Chap00PRA}.
Thus one can conclude that for the spin conversion in formaldehyde
quantum relaxation is an appropriate mechanism.   

\begin{figure}[htb]
\centerline{\psfig
{figure=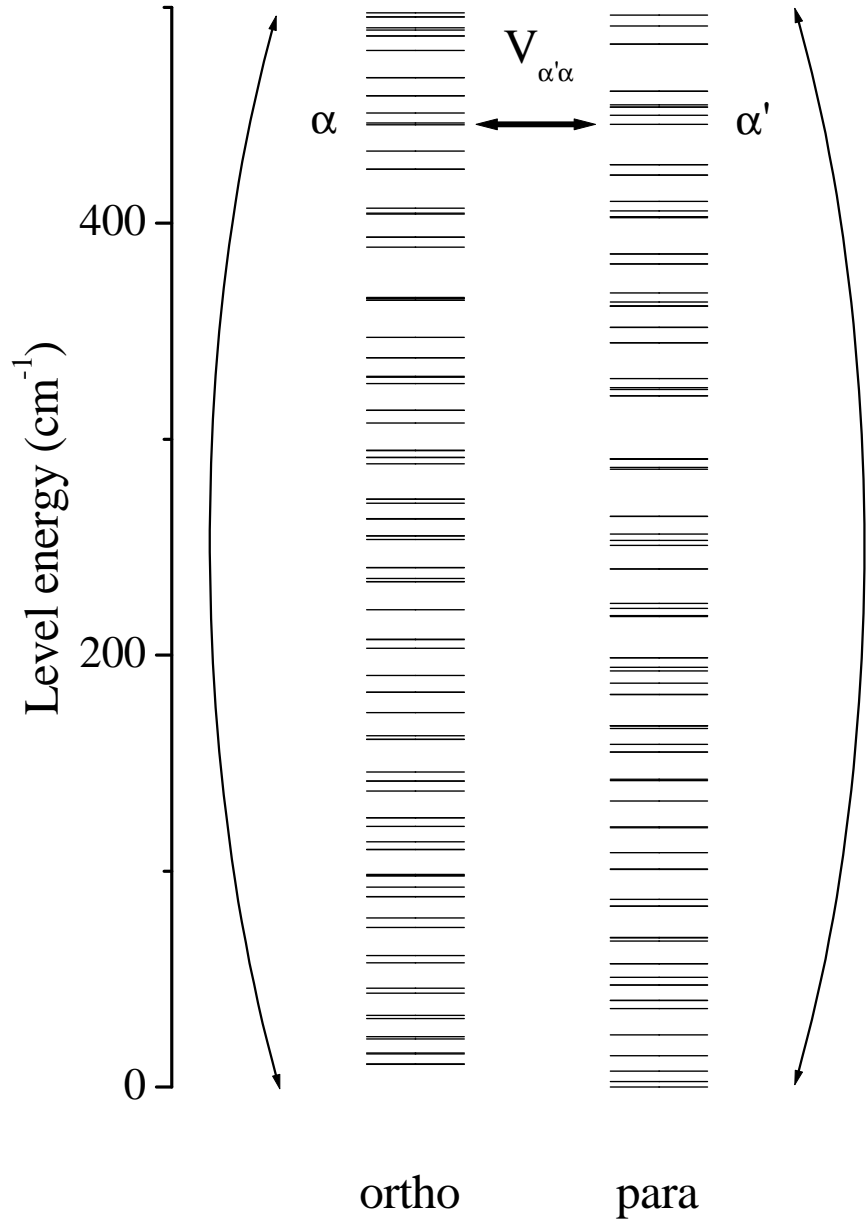,height=20cm}}
\vspace{0.5cm}
\caption{\sl Ortho and para states of formaldehyde (H$_2$CO). The levels are
calculated using the molecular parameters from Ref.~[7]. Bent lines indicate
transitions inside the ortho and para subspaces induced by collisions. The
level pair most important for the spin conversion in formaldehyde in shown to
be mixed by intramolecular perturbation $V$}
\label{fig1}
\end{figure}

The formaldehyde molecules have two nuclear spin isomers, ortho
(total spin of the two hydrogen nuclei, $I=1$) and para ($I=0$).
Each rotational state of the molecule can belong only to ortho,
or to para isomers. Thus the molecular states
are divided into two subspaces as it is shown in Fig.~\ref{fig1}.
Simple physical picture of spin conversion by quantum relaxation
is given elsewhere, e.g., \cite{Chap99ARPC}. Quantitative description 
of the process can be performed in the framework of the density
matrix formalism. The result of this description is as follows
\cite{Chap91PRA}. One has to split the molecular Hamiltonian
into two parts,
\begin{equation}
     \hat H = \hat H_0 +\hat V, 
\label{h}
\end{equation}
where the main part of the Hamiltonian, $\hat H_0$, has pure ortho 
and para states as the eigenstates; the perturbation $\hat V$
mixes the ortho and para states. If at initial instant the nonequilibrium
concentration of, say, ortho molecules, $\delta \rho _o(t=0)$, was created, 
the system will relax then exponentially, 
$\delta \rho _o(t)=\delta \rho_o(0)e^{-\gamma t}$, with the rate 
\begin{equation}
        \gamma = \sum_{a\in o, a'\in p}
        \frac{2 \Gamma F(a|a')}
        {\Gamma^2 + \omega^2_{aa'}}
        \left(W_o(\alpha) + W_p(\alpha')\right);\ \ 
        F(a|a') \equiv \sum_{\nu\in o, \nu'\in p}|V_{\alpha\alpha'}|^2.
\label{gamma}        
\end{equation}
where $\Gamma$ is the decay rate of the 
off-diagonal density matrix element $\rho_{\alpha\alpha'}\ 
(\alpha\in ortho;\ \ \alpha'\in para)$ assumed here to be equal 
for all ortho-para level pairs; $\omega_{aa'}$ is 
the gap between the 
states $a$ and $a'$; $W_o(\alpha)$ and $W_p(\alpha')$ are 
the Boltzmann factors of the corresponding states.
The sets of quantum numbers $\alpha\equiv\{a,\nu\}$ and 
$\alpha'\equiv\{a',\nu'\}$ consist of the degenerate quantum numbers
$\nu$, $\nu'$  and the quantum numbers $a$, $a'$ which 
determine the energy of the states. In Eq.~(\ref{gamma}) and further the ortho 
states will be denoted by unprimed characters, but para states by
primed characters. For the following it is 
convenient to introduce the {\it strength of mixing}, $F(a|a')$. 
In the definition of $F(a|a')$ in Eq.~(\ref{gamma}) the
summation is made over all degenerate states. 

The general model (\ref{gamma}) was tested comprehensively 
by conversion in symmetric tops ($^{13}$CH$_3$F, $^{12}$CH$_3$F)
and asymmetric top, $^{13}$CCH$_{4}$. It was proven that the ortho-para
mixing is performed in these molecules by intramolecular hyperfine
interactions. Thus, the spin isomer conversion
gives an alternative access to very weak intramolecular
forces which was investigated previously by high resolution spectroscopy,
e.g., Laser Stark Spectroscopy \cite{Duxbury85IRPC} and Microwave 
Fourier Transform Spectroscopy \cite{Bauder93}.   

To avoid confusion we stress that $\gamma$ from Eq.~(\ref{gamma}) gives
the equilibration rate in the system if one would measure the concentration
of ortho (or para) molecules. The authors \cite{Peters99CPL} introduced
the ortho-to-para ($k_1$) and para-to-ortho ($k_2$) rates which
are equal, respectively, to the first and second terms in the 
expression (\ref{gamma}) for $\gamma$. Thus $\gamma\equiv k_1+k_2$. 
Relation between the formaldehyde ortho and para partition functions
(see below) explains the equality $k_2=3k_1$ \cite{Peters99CPL}.
 
\section{Rotational states of formaldehyde}

\begin{figure}[htb]
\centerline{\psfig
{figure=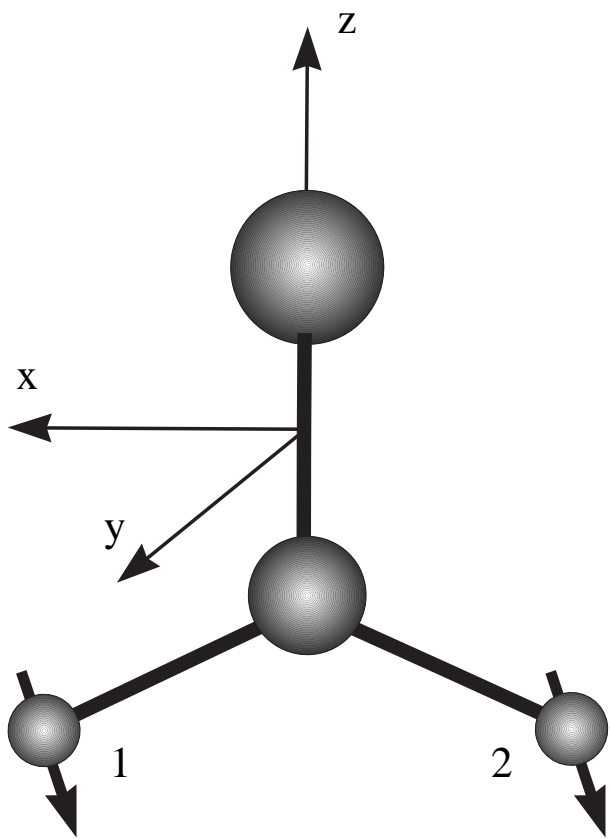,height=20cm}}
\vspace{0.5cm}
\caption{\sl Formaldehyde molecule, H$_2$CO, and orientation of the
molecular system of coordinates.}
\label{fig2}
\end{figure}

The formaldehyde molecule is a prolate, nearly symmetric top
having symmetry group C$_{2v}$. The characters of the group 
operations and its irreducible representations are given in the Table~1.
The molecular  structure and orientation of the molecular
system of coordinates are given in Fig.~\ref{fig2}.
The formaldehyde is a planar molecule having the following parameters 
in the ground state
$r_{CH}=1.1003\pm0.0005$~\AA, $r_{CO}=1.2031\pm0.0005$~\AA, 
and $\alpha_{HCO}=121.62\pm0.05^o$ \cite{Carter96JMS}. 

Rotational states of formaldehyde in the ground electronic
and vibrational state can be determined with high accuracy 
using the octic order Hamiltonian of Watson 
\cite{Watson77,Bocquet96JMS,Muller00JMS},
\begin{eqnarray}
\hat H_0 & = & \frac{1}{2}(B+C){\bf J}^2+(A-\frac{1}{2}(B+C))J^2_z -
       \Delta_J{\bf J}^4-\Delta_{JK}{\bf J}^2J^2_z -\Delta_K J^4_z \nonumber \\
      &&  + H_J{\bf J}^6 + H_{JK}{\bf J}^4J^2_z 
          + H_{KJ}{\bf J}^2J^4_z+H_KJ^6_z \nonumber \\
      &&  + L_{JJK}{\bf J}^6J^2_z + L_{JK}{\bf J}^4J^4_z
          + L_{KKJ}{\bf J}^2J^6_z + L_KJ^8_z \nonumber \\
      && + \frac{1}{4}(B-C)F_0-\delta_J{\bf J}^2F_0-
           \delta_K {}F_2 + h_J{\bf J}^4F_0
         + h_{JK}{\bf J}^2F_2 + h_KF_4 + l_{KJ}F_4,
\label{h0}
\end{eqnarray}
where ${\bf J}$, $J_x$, $J_y$, and $J_z$ are the molecular angular 
momentum operator and its projections on the molecular axes. 
The $B$, $C$, and $A$ are the parameters of a rigid top which 
characterize the rotation around $x$, $y$, and $z$ molecular axes, 
respectively (see Fig.~\ref{fig2}). The rest of parameters account 
for the centrifugal distortion effects \cite{Watson77}. 
In Eq.~(\ref{h0}) the notation was used
\begin{equation}
     F_n\equiv J^n_z(J^2_x-J^2_y) +  (J^2_x-J^2_y)J^n_z.
\label{f}
\end{equation}
We left in the Hamiltonian (\ref{h0}) only those terms for which
molecular parameters in \cite{Bocquet96JMS} were not set to zero.

It is convenient to diagonalise the Hamiltonian (\ref{h0}) in the
Wang basis \cite{Landau81},
\begin{eqnarray}
   |\alpha,p>  & = & \frac{1}{\sqrt{2}}
           \left[|\alpha> + (-1)^{J+K+p}|\overline\alpha>\right];
           \ \ 0<K\leq J,  \nonumber \\
   |\alpha_0,p>  & = & \frac{1 + (-1)^{J+p}}{2}|\alpha_0>;  \ \ K=0.      
\label{basis}
\end{eqnarray}
Here $p=0,1$; $|\alpha>$ are the symmetric-top rotational states; 
the sets of quantum numbers are $\alpha\equiv \{J,K,M\}$; 
$|\overline\alpha>\equiv \{J,-K,M\}$; $\alpha_0\equiv \{J,K=0,M\}$
where $J$, $K$, and $M$ are the quantum numbers of angular 
momentum and its projection on the molecular symmetry axis and 
on the laboratory quantization axis, respectively.  
Depending on the parity of $J$, $K$ and $p$, the states (\ref{basis}) 
generate 4 different irreducible representations of the
molecular symmetry group C$_{2v}$, as it is explained in the Table~1. 
In the following we will need the reduction of the 
matrix elements of full symmetric operator $\hat V$ in the basis 
$|\alpha,p>$ to the matrix elements of symmetric-top states
$|\alpha>$. This reduction reads
\begin{eqnarray}
 <\alpha,p|V|\alpha',p'> & = & 
             \delta_{p,p'}\left[ <\alpha|V|\alpha'> + 
             (-1)^{J'+p'}<\alpha|V|\overline{\alpha'}>\right];\ \ K'>0, \nonumber \\
 <\alpha,p|V|\alpha'_0,p> & = & \delta_{p,p'}
 \frac{1+(-1)^{J'+p'}}{\sqrt{2}} <\alpha|V|\alpha'_0>;\ \ K'=0.    
\label{rel}
\end{eqnarray} 

The molecular Hamiltonian, $\hat H_0$,
is full symmetric (symmetry A$_1$). Consequently, the matrix 
elements between the states of different
symmetry disappear. Thus diagonalization of the total Hamiltonian
in the basis of (\ref{basis}) is reduced to the diagonalization 
of four independent submatrices, each for the states of particular symmetry.
The rotational states of asymmetric top can be expanded 
over the basis states (\ref{basis}),
\begin{equation}
     |\beta,p> = \sum_K A_K|\alpha,p>,
\label{exp}
\end{equation}
where $A_K$ stands for the expansion coefficients.
The summation variable, $K$, is shown explicitly in (\ref{exp}),
although $A_K$ depends on other quantum numbers as well. All 
coefficients in the expansion (\ref{exp}) are real numbers because the 
Hamiltonian (\ref{h0}) is symmetric in the basis $|\alpha,p>$.

Complete description of the asymmetric-top quantum state needs 
indication of all
expansion coefficients, $A_K$, from (\ref{exp}) which is not
practical. There are a few schemes for abbreviate notations, see, e.g.,
\cite{Townes55}. We will use here the notations which are somewhat better
adopted to the consideration of the spin isomer problem in asymmetric tops 
\cite{Chap00PRA}. We will designate the rotational states of asymmetric top by 
indicating $p$, $J$ and prescribing the allowed $K$ values
to the eigen states keeping both in ascending order. For example,
the eigen state having $p=0,\ J=20$, the allowed $K$ in the
expansion (\ref{exp}) equal $K=0,2,4\dots$20 and being the third in ascending
order will be designated by ($p=0,\ J=20,\ {\cal K}=4$). Note the difference 
between the two characters $K$ and ${\cal K}$. It gives unambiguous notation 
of rotational states for each of the four species A$_1$, A$_2$ 
(${\cal K}$-even) and B$_1$, B$_2$ (${\cal K}$-odd). This classification 
becomes exact for a prolate symmetric top for which ${\cal K}=K$. 

Calculation of the level energies and wave functions of molecular
quantum states were performed in the paper numerically.
Accuracy of these calculations can be estimated
by comparing with the experimental rotational spectra in the ground
state of formaldehyde \cite{Bocquet96JMS}. This comparison shows
that the accuracy of the calculations for most rotational states 
is in the range of $10 - 100$~kHz. This should be sufficient for 
the investigation of the spin isomer conversion in formaldehyde. 

The two equivalent hydrogen nuclei in H$_2$CO have spin 1/2.
It implies that the total wave function (product of spin and spatial 
wave functions) is of symmetry B$_1$.  
Spin states of the two hydrogen nuclei can be either 
triplet (ortho, $I=1$, symmetry A$_1$), or singlet (para,
$I=0$, symmetry B$_1$). In order to have the total wave function
of symmetry B$_{1}$, the ortho molecules should have the spatial
wave function of symmetry B$_{1}$, but the para molecules should have 
the spatial wave function of symmetry A$_{1}$. Consequently, in
the ground electronic and vibrational state the 
rotational states A$_1$ and B$_1$ are only positive (even in parity), 
but the states A$_2$ and B$_2$ are only negative (odd in parity). 

The ortho states of the formaldehyde molecule can be presented as
\begin{equation}
     |\mu> = |\beta,p>|I=1,\sigma>; \ \ {\cal K}-{\text{odd}}.
\label{ortho}
\end{equation}
where $\sigma$ is the projection of the nuclear spin {\bf I}
on the laboratory quantization axis. The para states can be presented as
\begin{equation}
     |\mu'> = |\beta',p'>|I'=0>;\ \ {\cal K}'-{\text{even}}.
\label{para}
\end{equation}

The Boltzmann factors $W_o(\alpha)$ and $W_p(\alpha')$ in 
Eq.~(\ref{gamma}) determine the population of 
the  states $\alpha$ and $\alpha'$ in the ortho and para families,
\begin{equation}
     \rho_{\alpha} = \rho_o W_o(\alpha);\ \ 
     \rho_{\alpha'} = \rho_p W_p(\alpha'),
\label{Bolt}
\end{equation}
where $\rho_o$ and $\rho_p$ are the total densities of ortho 
and para molecules, respectively. 
The partition functions for ortho and para molecules at room 
temperature (T=300~K) are found to be equal to
\begin{equation}
     Z_{ortho}=2.16\cdot10^3;\ \ Z_{para}=721.
\label{z}
\end{equation}
In the calculation of these partition functions the energies of the 
rotational states were determined numerically using the 
Hamiltonian (\ref{h0}). The degeneracy over $M$, $\sigma$, 
as well as the restrictions imposed by the 
quantum statistics were taken into account. 

\section{Mixing of the ortho and para states}

There are two known intramolecular perturbations able to mix 
ortho and para states in polyatomic molecules. The first one is
the spin-spin interaction between the molecular nuclei. This 
interaction has simple form in formaldehyde and can be 
expressed as \cite{Landau81}
\begin{eqnarray}
     \hat V_{SS} &\ =\ & 
     P_{12}\hat{\bf I}^{(1)}\hat{\bf I}^{(2)}{\ {}^\bullet_\bullet\ }
     {\bf T}\ ; \nonumber \\
     T_{ij}     &\ =\ &
     \delta _{ij}-3\delta_{i,x}\delta_{j,x}\ ;\ \ 
     P_{12}=\mu^2_p/r^3I^{(1)}I^{(2)}h\ ,
\label{vss}     
\end{eqnarray}
where $\hat{\bf I}^{(1)}$ and $\hat{\bf I}^{(2)}$ are the spin 
operators of the hydrogen nuclei 1 and 2; $\mu_p$ is magnetic moment of
proton; $i$ and $j$ are the Cartesian indices. 
The second rank tensor $\hat{\bf I}^{(1)}\hat{\bf I}^{(2)}$ acts
on spin variables. The second rank tensor {\bf T} 
represents a spatial part of the spin-spin 
interaction. One can deduce from the angular momentum
algebra that all matrix elements of the perturbation 
$\hat V_{SS}$ between ortho and para states of formaldehyde 
are vanishing. This is because one 
cannot draw a triangle having the sides 2, 1, and 0, which are
the rank of the tensor $\hat{\bf I}^{(1)}\hat{\bf I}^{(2)}$, the
total spin of ortho, and para states, respectively.

The second intramolecular perturbation which should be
considered is the spin-rotation coupling
between spins of hydrogen nuclei and molecular rotation. 
The spin-rotation coupling can be 
presented in general as \cite{Gunther-Mohr54PR,Townes55,Ilisca98PRA}
\begin{equation}
     \hat V_{SR} \equiv \sum_n \hat V^{(n)}_{SR} = \frac{1}{2}
     \left(\sum_n \hat {\bf I}^{(n)}\bullet {\bf C}^{(n)} 
     \bullet \hat{\bf J} + h.c.\right);\ \ n=1,2.
\label{vsr}
\end{equation}
Here {\bf C} is the spin-rotation tensor; $\hat{\bf J}$ 
is the angular momentum operator. Index $n$ in (\ref{vsr}) refers only 
to the hydrogen nuclei because we are interested now in the perturbation
able to mix ortho and para states. 

Calculation of the spin-rotational tensor, {\bf C}, is a complicated  problem.
Further, a few simplifications will be made. First, we neglect small 
contribution to the tensor {\bf C} due to the electric fields
at the position of protons\cite{Bahloul98JPB}. The remaining part of 
the tensor {\bf C} originates from the magnetic fields 
produced by the electrical currents in
the molecule. One can split the tensor {\bf C} into two terms,
\begin{equation}
{\bf C}={}^e{\bf C}+{}^n{\bf C},     
\label{split}
\end{equation}
where ${}^e{\bf C}$ and ${}^n{\bf C}$ are due to the
electron and nuclear currents in the molecule, respectively. The
electron part, ${}^e{\bf C}$, appears as an average over electron state,
which has nonzero value only in the second order perturbation theory.
Its calculation needs the knowledge of electron excited states and
thus is rather difficult to perform. On the other hand, the nuclear part,
${}^n{\bf C}$, has nonzero value for the ground electron state and
can be easily estimated. In the following we will neglect the electron part,
${}^e{\bf C}$, and will use the estimation {\bf C}$\simeq{}^n{\bf C}$. This
can be considered as an upper limit for {\bf C} because molecular electrons
are following the rotation of nuclear frame and thus compensate 
partially the magnetic field produced by nuclei. 

The spin-rotation tensor ${}^n{\bf C}$ can be presented (in Hz) 
as \cite{Bahloul98JPB} 
\begin{eqnarray}
     {\text{C}}^{(n)} & = &{\sum_{k\neq n}} b_k\left[ 
          ({\bf r}_k\bullet {\bf R}_k){\bf 1} - 
           {\bf r}_k {\bf R}_k\right]\bullet {\bf B}; \nonumber \\
       b_k &=& 2 \mu_p q_k\big/ c \hbar R^3_k \ ,
\label{nc}
\end{eqnarray}
where {\bf R}$_k$ is the radial vector from the proton H$^{(n)}$
to the  charge $k$; {\bf r}$_k$ is the radial from the center of mass to 
the particle $k$; $q_k$ are the nuclei' charges; {\bf B} is the inverse 
matrix of inertia moment. {\bf B} is a diagonal matrix having the 
elements $B_{xx}= 68.0$~GHz, $B_{yy}=77.67$~GHz, and $B_{zz}=563.9$~GHz.
Index $k$ runs here over all nuclei in the molecule except the
proton $n$. 

Using the symmetry operation C$_2$ one can prove the equality of 
the two matrix elements, $<\mu|V^{(1)}_{SR}|\mu'>=<\mu|V^{(2)}_{SR}|\mu'>$.
Thus for the evaluation of the spin-rotation coupling in 
formaldehyde it is sufficient to calculate one matrix element, e.g.,
$<\mu|V^{(1)}_{SR}|\mu'>$. We write this matrix element using an expansion
over symmetric-top states (\ref{exp}),
\begin{equation}
     <\mu|V^{(1)}_{SR}|\mu'> = <I=1,\sigma|
     \left[\sum_{K,K'}A_K A'_{K'} 
     <\alpha,p|V^{(1)}_{SR}|\alpha',p'>\right]|I'=0>.    
\label{exp1}
\end{equation}
This expression and (\ref{rel}) reduce the calculation 
of the spin-rotation matrix 
elements of asymmetric tops to the calculation of symmetric-top 
matrix elements. Solution for the latter can be found 
in \cite{Guskov95JETP,Ilisca98PRA,Guskov99JPB}, which allows to 
express the strength of mixing in formaldehyde by $\hat V_{SR}$ as
\begin{eqnarray}
  F_{SR}(a'|a)& = & \frac{1}{4}(2J'+1)(2J+1)\Bigg|
                   \sum_{K>0,q}A_{K+q}A'_K 
                    \Phi(J,K'+q|J',K')  \nonumber \\
          &  & +\frac{1+(-1)^{J'+p'}}{\sqrt{2}}A_1A'_0
                \Phi(J,1|J',0)\Bigg|^2 .    
\label{fsr}        
\end{eqnarray} 
Here $q=\pm1$; In (\ref{fsr}) the notation was used 
\begin{eqnarray}
    \Phi(J,K|J',K') & = & \sum_l \sqrt{2l+1}\, {\cal C}_{l,q}
                  \left(\begin{array}{ccc}    
                      J' & l & J \\            
                     -K' & q & K               
                  \end{array}\right)  \times   \nonumber \\           
       &  &  \left[ y(J)(-1)^l \left\{\begin{array}{rcr}    
                                     J'& J & l \\            
                                     1 & 1 & J               
                               \end{array}\right\} + 
                  y(J') \left\{\begin{array}{rcr}    
                               J & J'& l \\            
                               1 & 1 & J'               
                               \end{array}\right\} \right],
\label{fjk}
\end{eqnarray} 
where (:~:~:) stands for the 3j-symbol; 
\{:~:~:\} stands for the 6j-symbol; $y(J)=\sqrt{J(J+1)(2J+1)}$;  
${\cal C}_{l,q}$ are the spherical components of the spin-rotation
tensor of the rank $l$ ($l=1,2$) for the first proton calculated in 
the molecular frame. ${\cal C}_{l,q}$ can be determined using 
Eq.~(\ref{nc}). For the formaldehyde molecular structure from the 
Ref.~\cite{Carter96JMS} and bare nuclei' charges these components are
\begin{equation}
     {\cal C}_{2,1}=3.39~{\text{kHz}};\ \   
             {\cal C}_{1,1}=-3.39~{\text{kHz}}.
\label{c}
\end{equation}
We stress that these values give an upper limit to the {\bf C}-tensor.

The selection rules for the ortho-para mixing by spin-rotation 
perturbation in formaldehyde read
\begin{equation}
     \Delta p = 0;\ \ |\Delta J| \leq 1.
\label{selSR}
\end{equation}
Parity of ${\cal K}'$ and ${\cal K}$ is opposite. 

\section{Conversion rates}

For the calculations of the isomer conversion rate one needs the value
of the ortho-para decoherence rate, $\Gamma$, see Eq.~(\ref{gamma}). 
Experimental determination of this parameter can be based on the level-crossing
resonances in spin conversion. Such measurements were performed 
so far only for the $^{13}$CH$_3$F spin conversion and gave
$\Gamma/P\simeq2\cdot10^8$~s$^{-1}$/Torr (see the discussion
in \cite{Chap99ARPC}). Estimation of $\Gamma$ can be done 
using the pressure line broadening data. For polar molecules the line 
broadening is on the order of $\sim10^8$~s$^{-1}$/Torr. Further, we
will assume the decoherence rate being equal to
\begin{equation}
     \Gamma/P=1\cdot10^8~{\text{s}}^{-1}{\text{/Torr}}.
\label{G}
\end{equation}
The same value of $\Gamma$ was used in \cite{Peters99CPL}.

\begin{figure}[htb]
\centerline{\psfig
{figure=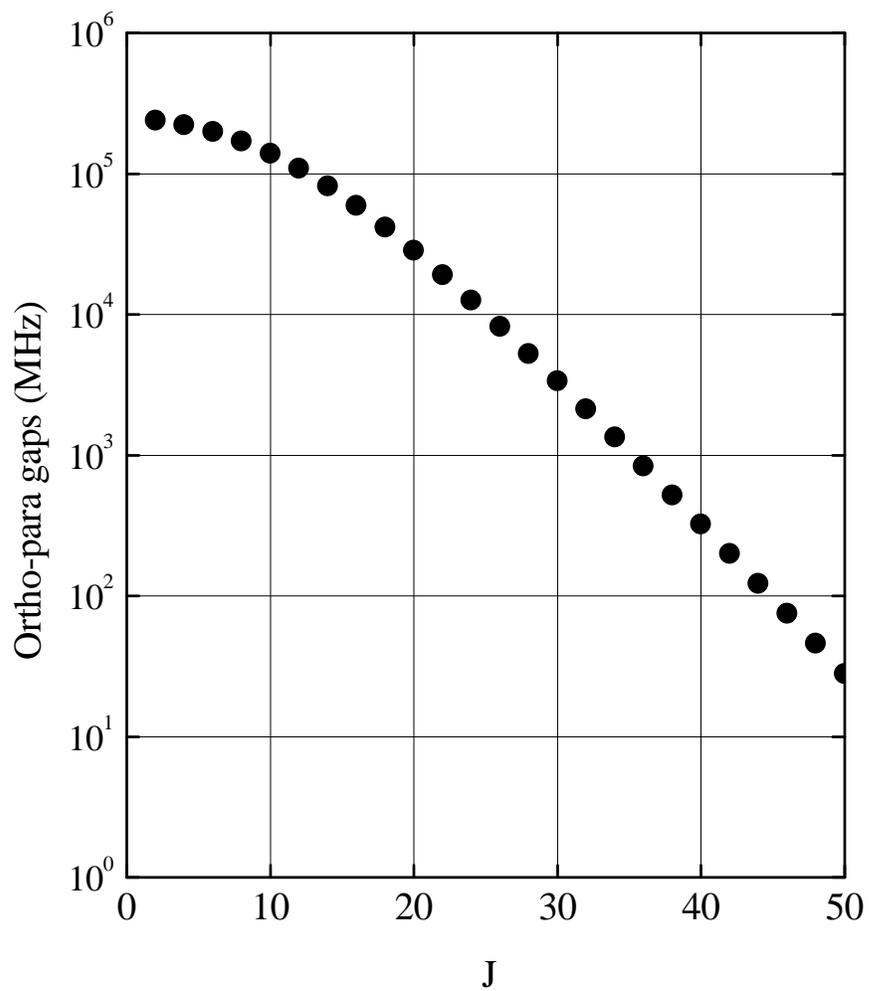,height=20cm}}
\vspace{0.5cm}
\caption{\sl Frequency gaps in the egular sequence of close ortho
($p=0$,$J$,${\cal K}=1$) and para $p'=0$,$J$,${\cal K}'=0$) states in 
formaldehyde.}
\label{fig3}
\end{figure}

It is clear from Eq.~(\ref{gamma}) that only close
ortho-para level pairs can contribute significantly to the spin
conversion. Formaldehyde is a light molecule having large level 
spacing. The average density of levels in 
the range $0-1000$~cm$^{-1}$ is low, 1 level per each 5~cm$^{-1}$.
As was pointed out already in \cite{Curl67JCP}, there are regular
and ``accidental'' ortho-para resonances. An example of regular resonances   
$(p=0,J,{\cal K}=1) - (p'=0,J,{\cal K}'=0)$ is shown in Fig.~\ref{fig3}.
The ortho-para gaps in this sequence of states goes rapidly down as
$\sim\exp(-0.23J)$.  Analogous phenomenon of collapsing ortho and para states
exists also in ethylene where decrease of gaps is even faster
\cite{Chap00PRA}. 

Calculated spin conversion rates are 
given in the Table~2. The total conversion rate combines
contributions from all ortho-para level pair having $J$ up to
40 and $|\omega|<40$~GHz. Thus the spin conversion rate in
formaldehyde is
\begin{equation}
     \gamma/P = 1.4\cdot10^{-4}~{\text{s}}^{-1}/{\text{Torr}}.
\label{total}
\end{equation}

\begin{figure}[htb]
\centerline{\psfig
{figure=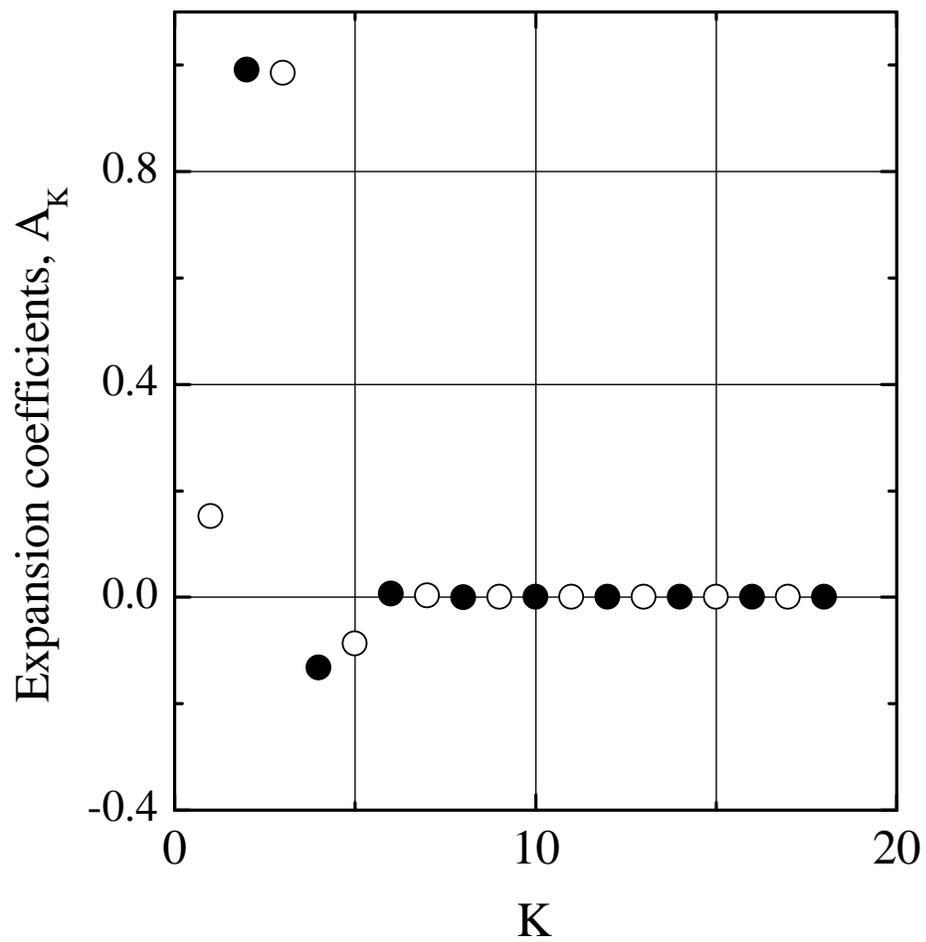,height=20cm}}
\vspace{0.5cm}
\caption{\sl The expansion coefficients, $A_K$, for the states 
most important for the spin conversion in H$_2$CO. 
(o)--ortho state ($1,J=17,{\cal K}=3$); 
($\bullet$)--para state ($1,J'=18,{\cal K}'=2$).}
\label{fig4}
\end{figure}

One can conclude from the data presented in Table~2 that
the close sequence of states from Fig.~\ref{fig3} does not
contribute significantly to the conversion. The same effect was found in
the ethylene conversion \cite{Chap00PRA}. Another observation 
from the Table~2 is that there are no close ortho-para level
pairs in formaldehyde which can be mixed by the spin-rotation coupling.
The most important ortho-para level pair which contributes
more than 50\% to the total rate is the pair (1,18,2)--(1,17,3). 
It has the energy gap $\simeq6$~GHz. The 
expansion coefficients, $A_K$, for these states are presented in
Fig.~\ref{fig4}. One can see from these data that the wave functions
of these states are rather close to the symmetric-top case which
would have just one term in the expansion (\ref{exp}).

\section{Discussion and Conclusions}  

Calculated value of the conversion rate in formaldehyde was found to 
be almost 10 times smaller than the experimental one \cite{Peters99CPL}.
There are two main uncertainties in the present calculations
which both can be defined more accurately by future experiments. 
The first uncertainty comes from the decoherence rate, $\Gamma$.
This parameter can be determined by careful study of the
pressure broadening of rotational lines in the formaldehyde
ground state. The second uncertainty originates from poor knowledge 
of the spin-rotation coupling in formaldehyde. The spin-rotation
coupling in H$_2$CO can be investigated using high resolution 
spectroscopy methods, e.g., Laser Stark Spectroscopy \cite{Duxbury85IRPC} 
and Microwave Fourier Transform Spectroscopy \cite{Bauder93} which were
proven to be efficient for the investigation of hyperfine interactions
in molecules.

In general, it is difficult  to verify the mechanism of
spin conversion in formaldehyde by comparing single values of the theoretical
and experimental conversion rates, which depends in theory on a number of parameters 
and in experiment can be resulted from a few effects, e.g., chemical reactions.    
It is more appropriate to compare the dependencies predicted by the theory
with the experimental dependencies. First, the model predicts rather 
strong dependence of the conversion rate on the type of buffer gas. 
By varying the collision partner one can change the decoherence 
rate, $\Gamma$, by nearly one order of magnitude. In the same proportion
the spin conversion rate should be changed if conversion is governed 
by quantum relaxation. It is alarming that
the authors \cite{Peters99CPL} observed very small change of $\gamma$ by 
adding the argon gas up to the pressure of 760~Torr. 

\begin{figure}[htb]
\centerline{\psfig
{figure=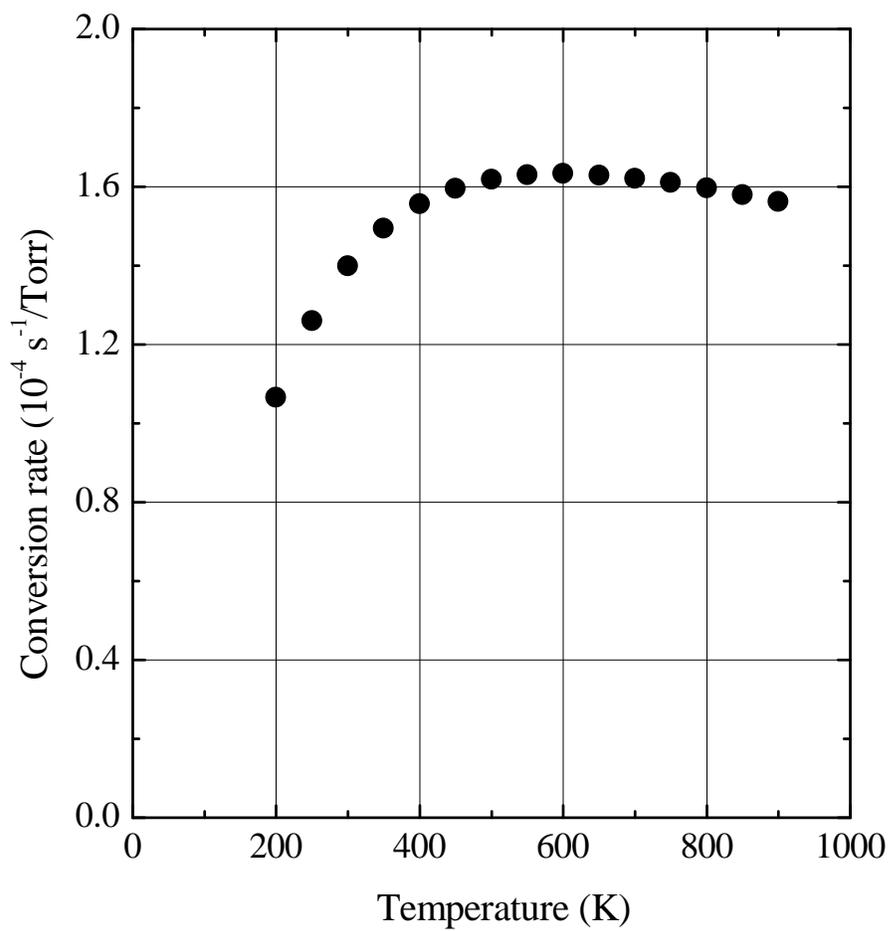,height=20cm}}
\vspace{0.5cm}
\caption{\sl Temperature dependence of the total spin conversion rate in
formaldehyde.}
\label{fig5}
\end{figure}

Another experimental verification of the conversion mechanism  
could be the investigation of temperature dependence.
The theoretical temperature dependence of the spin conversion
rate in formaldehyde is shown in the Fig.~\ref{fig5}. The 
calculation were done under an assumption that the decoherence
rate, $\Gamma$, is temperature independent. Such assumption is 
supported by the slow temperature
dependence of $\Gamma$ observed in the case of spin conversion
in $^{13}$CH$_3$F \cite{Chap99ARPC}. The theoretical model predicts 
rather weak influence of the gas temperature on the 
formaldehyde conversion
rate, $\gamma$, in the wide range of temperatures, $T=200-900$~K. 

In conclusion, we have developed theoretical model of the spin conversion
in formaldehyde. Although this model has the same basic concepts as
the model developed in \cite{Curl67JCP}, the key parameters of the 
new model are more precise. First of all, it refers to the
molecular level energies and the wave functions. 

We have analysed the relation between the theoretical model of the
formaldehyde spin conversion and the experiment \cite{Peters99CPL}. 
The theoretical model in its present form gives 10 times smaller conversion
rate than the rate measured in \cite{Peters99CPL}. Two types of experiments 
have been proposed which can help to resolve the puzzle.

\newpage
Table~1. The character table for the C$_{2v}$ symmetry group and 
the classification of the basis states (\ref{basis}).

\vspace{1cm}
\begin{tabular}{|c|cccc||ccc||c|}
\hline
       &E&C$_{2}$&$\sigma_v$&$\sigma_v'$&$K$-even & $K$=0     & $K$-odd & Sign \\
\hline                                                                                      
A$_{1}$&1&    1  &     1    &   1       &$p$=0    &$J,p$-even,&--       &  +   \\
B$_{2}$&1&   -1  &    -1    &   1       &--       &--         &$p$=1    &  -   \\
A$_{2}$&1&    1  &    -1    &  -1       &$p$=1    &$J,p$-odd, &--       &  -   \\
B$_{1}$&1&   -1  &     1    &  -1       &--       &--         &$p$=0    &  +   \\
\hline
\end{tabular}

\newpage
Table~2. The most important ortho-para levels and 
their contributions to the spin conversion in formaldehyde.

\vspace{1cm}
\begin{tabular}{|ccccc|}
\hline
          Level pair   & Energy & $\omega/2\pi$ & F$_{SR}$ & $\gamma/P$ \\
 $p',J',{\cal K}'$-$p,J,{\cal K}$&(cm$^{-1}$)&(MHz) & ($10^{-2}$~MHz$^2$)         
                                                           &(10$^{-5}$~s$^{-1}$/Torr)\\
\hline
  1,18,2--1,17,3 & 445.87 & -6162 & 6.59  & 7.56  \\
   0,8,2--0,9,1  & 120.93 &  9252 & 0.98  & 2.38  \\
  1,21,4--1,22,3 & 692.52 & -10379& 12.5  & 1.55  \\
  0,13,2--0,14,1 & 253.82 &  17211& 2.45  & 0.91  \\
  1,12,2--1,13,1 & 221.18 & -19305& 2.01  & 0.69  \\
\hline
 Total rate &  & & & 14.0 \\
\hline
\end{tabular}

\end{document}